\documentclass[showpacs,preprintnumbers,amsmath,amssymb]{revtex4}

\usepackage{graphicx}
\usepackage{bm}
\usepackage{epstopdf}
\usepackage{float}
\usepackage{dcolumn}
\usepackage{bm}
\usepackage{mathrsfs}
\usepackage{amsmath}
\usepackage{color}

\begin{document}

\title{Quantum-Mechanical Corrections to the Schwarzschild Black Hole Metric}
\author{P. Bargue\~no}
\email{p.bargueno@uniandes.edu.co}
\author{S. Bravo Medina}
\email{s.bravo58@uniandes.edu.co }
\author{M. Nowakowski}
\email{mnowakos@uniandes.edu.co}
\affiliation{%
Departamento de Fisica,\\ Universidad de los Andes, Cra.1E
No.18A-10, Bogota, Colombia
}
\author{D. Batic}
\email{davide.batic@uwimona.edu.jm}
\affiliation{%
Department of Mathematics, University of West Indies,
Kingston 6, Jamaica \\
and \\
Department of Mathematics, The Petroleum Institute, Abu Dhabi, UAE}

\date{\today}

\begin{abstract}
Motivated by quantum mechanical corrections to the Newtonian potential,
which can be translated into an $\hbar$-correction to the $g_{00}$ component of the Schwarzschild metric, 
we construct a quantum mechanically corrected metric assuming
$-g_{00}=g^{rr}$. We 
show how the Bekenstein black hole entropy $S$ receives its logarithmic
contribution provided the quantum mechanical corrections to the 
metric are negative. 
In this case the standard horizon at the Schwarzschild radius $r_S$ 
increases by small terms proportional to $\hbar$ and a remnant of the
order of Planck mass emerges. 
We contrast these results with a positive correction to the metric 
which, apart from a corrected Schwarzschild horizon,
leads to a new purely quantum mechanical
horizon. 
\end{abstract}

\pacs{Valid PACS appear here}
\maketitle
\section{Introduction}
The full theory of Quantum Gravity is one of the last unsolved challenges in fundamental science and is still eluding us. 
Nevertheless, some effects of Quantum Theory do enter the context of the theory of Quantum Gravity and can
be handled in a rigorous way without the knowledge of the full fledged theory. 
Such is the case of the Hawking radiation \cite{Hawking} and/or the Unruh effect \cite{Unruh}. 
Apart from these paradigms there are some other interesting quantum effects related to gravity like
the absence of stable orbits of fermions around a black hole 
\cite{Fermions}, the quantum correction to the Bekenstein
entropy $S$ of black holes and of other black objects using different approaches to Quantum Gravity 
\cite{Kaul2000,Medved2004,Camelia2004,Meissner2004,Das2002,Domagala2004,Chatterjee2004,
Akbar2004,Myung2004,Chatterjee2005,Adler2001,Camelia2006,Majumder2011,
Majumder2013,Gorji2014,Faizal1,Faizal2,Faizal3,Faizal4,Carlip2000,Ghosh2005,Fursaev1995,Nojiri2003,Kaul2012,Hod2004,Hai-Xia2007,Tawfik2015,Mukherji2002,Wang2008,Banerjee2008} and   
the quantum correction to the Newtonian potential or metrics \cite{Donoghue1994-1,Donoghue1994,HamberLiu,Muzinich,Akhundov,KK,PRD2003a,PRD2003,
Ross-Holstein,Kirilin,Haranas-Mioc,Donoghue2012,BDH2015,Rovelli,LivingReviews,Akhundov2,Woodard,Radkowski,CapperDuff} (for some applications of the
new corrections see \cite{Burns, Haranas, Faller}).
Indeed, the results regarding the corrected Newtonian potential $\Phi$ spread over a period of the last forty five years 
starting with the early seventies
whereas the corrections to $S$ are a relative new undertaking. Whichever model one uses it turns out that
$S$ receives corrections proportional to the logarithm of the black hole area and, in some models, also proportional
to the square root of this quantity.
This is also the finding of our approach starting from a different context.  
We will make a connection between the $\hbar$-corrected metric and the quantum mechanical corrections to the entropy.
We will construct our quantum mechanically corrected metric by demanding (i) that it reproduces the $\hbar$ corrected
Newtonian limit, (ii) that it reproduces the standard result for the entropy of black hole including, in addition,
the $\hbar$ corrections which are similar to results established elsewhere and (iii) that it passes some consistency
checks regarding the geodesic motion of a test particle moving in this metric. Point (i) which has to do
with weak gravity, can
be easily accommodated by invoking the classical connection between the $g_{00}$ metric component and the Newtonian
potential.  The second point requires the determination of the horizons and probes
into the strong regime of gravity. In principle, we cannot infer the strong gravity effects from
results zeroing around the weak regime as is the case of quantum corrections to the Newtonian
potential. However, we let ourselves be guided by the fact that in the most radially symmetric metrics
the time component is inverse of the radial one. We will take over this fact to
the quantum mechanically corrected metric and show that this step is sufficient
to derive the quantum correction to the Bekenstein entropy. Strictly speaking,
this step is justified a posteriori as it enables us to obtain the right result. Finally,
we check how the equation of motion of a test particle gets affected by the quantum corrections.
If overall the new metric is in accordance with observational facts including the classical tests
of General Relativity we can then consider such a result as consistent.
 
We note that
we arrive at the standard results for the black hole entropy (logarithmic and polynomial corrections) obtained in different ways elsewhere.
This
gives us some confidence about the $\hbar$-corrections to the metric and the way we handle the calculation. Again, our results
show that it is not necessary to invoke the full machinery of a particular Quantum Gravity theory to
derive a valid quantum mechanical result in gravity. Indeed, the quantum corrections to $\Phi$ have been
obtained by treating gravity as an effective field theory, which is a conventional approach.

The paper is organized as follows. In section II we motivate the metric and present its full form.
We give the first insight into the horizons connected with the metric. Section III is devoted to the
thermodynamics of the black hole governed by the quantum-mechanically corrected metric. Here we calculate
the corrections to the Bekenstein black hole entropy. In the subsequent section we show by means of the
heat capacity that a black hole remnant emerges. To illuminate the role of the sign of the
quantum corrections the next section treats a hypothetical case. This is followed by a section in which we compare our results
with results obtained in literature. Section VII discusses the geodesic equation of motion
resulting from the new metric. This serves as a consistency check to show that no unwanted features will appear
in the motion of a test particle. In the last section we draw our conclusions.  
\section{$\hbar$-correction to the metric}
In the field theoretical language of an effective field theory of gravity the Fourier transform of an elastic
scattering amplitude gives the potential in $r$. The one loop correction being always proportional to $\hbar$
represents then (after the Fourier transform) the quantum mechanical correction to the potential
under discussion. 
The terms considered in such a calculation are non-analytic terms which come from
the propagation of two or more massless particles (gravitons) in the Feynman diagrams. These terms
dominate over the analytic contributions in the low energy limit of the effective theory \cite{PRD2003a}.
They are of the type $1/\sqrt{-q^2}$ and $\ln(q^2)$ where $q^2$ is the momentum transfer. After the Fourier transform the first term
emerges as proportional to $1/r^2$ whereas the second one gives $1/r^3$ terms.
The full result is often written in the form 
\begin{equation} \label{N1}
\Phi(r)=-\frac{G M_{1}M_{2}}{r}\left[ 1 +\lambda \frac{G(M_{1}+M_{2})}{rc^{2}}-\gamma \frac{G \hbar}{r^{2}c^{3}}+\ldots \right]
\end{equation}
where the $\lambda$ and $\gamma$ are parameters which take different values 
depending on the framework under consideration.
Partly, we can attribute the reason for the discrepancies in $\lambda$ to the 
precise coordinate definition used in the calculation \cite{LivingReviews}. 
The question about the ambiguity of this potential, due to the lack of clarity on the coordinates, has
also been put forward in some related articles \cite{PRD2003a,LivingReviews,Gauge}.
It is argued that a redefinition $r\rightarrow r'=r(1+aGM/r)$ would change the 
parameter $\lambda$ without affecting the observables. The general consensus is
that we can write the corrected potential as \cite{LivingReviews, Akhundov2}
\begin{equation} \label{N2}
\Phi(r)=-\frac{GM_{1}M_{2}}{r}\left[1-\gamma \frac{G\hbar}{r^{2}c^{3}}+\ldots \right]
\end{equation}
taking $r$ to be the distance between two objects with masses $M_1$ and $M_2$, namely the static Schwarzschild
$r$ \cite{KK,Duff} and not the gauges considered in other references. 
Hence, we note that the 
terms proportional to $\lambda$ in
(\ref{N1}) are not additional relativistic corrections, but artifacts of the different choice
of coordinates.
The aforementioned re-parametrization freedom still cannot account for all the
discrepancies of the different $\gamma$'s found in the literature. A
number of errors have been identified \cite{PRD2003a,Donoghue1994,DonoghueReview}, but
it is not clear if this accounts for all the different values available.
It is therefore fair to list some of the results (see Table \ref{table:nonlin}).
In Table I we have collected the different values for $\gamma$ which also vary in sign (we will see
that the sign plays the most important role in the phenomenology derived from these corrections).
It is also worth noting that the value in \cite{HamberLiu} differs from the result in \cite{BDH2015} not because
of an error which occurred in the calculation, but due to a deliberate choice
of the amplitude: \cite{HamberLiu} considers only the one-particle reducible amplitude in contrast to. e.g. \cite{BDH2015} where the full
amplitude is taken into account. 

\begin{table}
\caption{Different values of $\gamma$ found in the literature.}
\begin{center}
\begin{tabular}{ |c|c| } 
 \hline
 \textbf{(Year) Reference} & $\gamma$ 
 \\
 \hline 
 (1994) \cite{Donoghue1994} &   \parbox[c][4ex]{6ex}{\centering $\frac{127}{30\pi^{2}}$}\\ 
 \hline 
 (1995) \cite{HamberLiu} & \parbox[c][4ex]{6ex}{\centering $\frac{122}{15\pi}$}\\
 \hline
 (1995) \cite{Muzinich} & \parbox[c][4ex]{6ex}{\centering $-\frac{17}{20\pi}$}\\
 \hline
 (1998) \cite{Akhundov} & \parbox[c][4ex]{6ex}{\centering $\frac{107}{10\pi^{2}}$}\\
 \hline
 (2002) \cite{KK} & \parbox[c][4ex]{6ex}{\centering $-\frac{121}{10\pi}$}\\
 \hline
 (2003) \cite{PRD2003a} & \parbox[c][4ex]{6ex}{\centering $-\frac{41}{10\pi}$}\\
 \hline
 (2003) \cite{PRD2003} & \parbox[c][4ex]{6ex}{\centering $-\frac{167}{30\pi}$}\\
 \hline
 (2007) \cite{Ross-Holstein} & \parbox[c][4ex]{6ex}{\centering $-\frac{41}{10}$ } \\ 
 \hline
(2007) \cite{Kirilin} & \parbox[c][4ex]{6ex}{\centering $\frac{107}{30\pi}$ } \\ 
 \hline
 (2002) \cite{Haranas-Mioc} & \parbox[c][4ex]{6ex}{\centering $\frac{122}{15\pi}$}\\
 \hline
(2012) \cite{Donoghue2012} & \parbox[c][4ex]{6ex}{\centering $-\frac{41}{10\pi}$}\\
 \hline
 (2015) \cite{BDH2015} & \parbox[c][4ex]{6ex}{\centering $-\frac{41}{10}$}\\
 \hline
\end{tabular}
\end{center} 
\label{table:nonlin} 
\end{table}
Given the history of the subject, our approach here will be to take the latest value (see, however, \cite{Woodard}), as the correct one, i.e, $\gamma=-41/10$.
In any case, it is not so much the numerical value of $\gamma$ which affects the conclusions, but the sign.
By requiring absence of negative norm states it has been argued in \cite{Dvali1, Dvali3} that modified gravity cannot become weaker
than the pure Einstein gravity. This would hint towards a negative $\gamma$. We can make the argument more conclusive by
using the lower bound on the dimensionless weak coupling $\alpha_g$ which according to \cite{Dvali1, Dvali2} reads $\alpha_g(p) \ge p^2/M_p$ where
$M_p=1/\sqrt{16\pi G}$ (reduced Planck mass) and $p$ is a typical momentum in the process which we interpret here as the momentum transfer
$q$. Since $q \sim 1/r$ we can tentatively estimate that $\alpha_g(r) \ge 1/(r^2M_p)$. In pure gravity the dimensionless quantity $ \alpha_g(r)$
is simply $G/r^2$ and the Newtonian potential reads $\Phi=-\alpha_g(r)M_1M_2r$. In the quantum corrected gravity we have
$\alpha_g(r)=(G/r^2)[1- \gamma (l_p/r)^2]$  ($l_p$ denotes the Planck length defined by $l_p=G\hbar/c^3$) 
and the inequality from above tells us that $-\gamma (l_p/r)^2 \ge 0$ which implies a negative
$\gamma$.

At the level of the amplitude, unitarity is violated at the energy $E^2_{\rm CM}=20(GN)^{-1}$ where $N$ is a sum of scalars, fermions and vector
particles \cite{Han}. This would indicate that the validity of the approach breaks down as we are approaching the Planck scale. It is argued
in \cite{Aydemir} that the violation of unitarity at the tree level is ``self-healing'' through iteration of vacuum polarization, but the effect on
the correction to the Newtonian potential remains open. Therefore, we will mention it explicitly if our conclusions touch the
Planck region. We kept the presentation on the $\hbar$ corrections brief since excellent reviews on the subject already exist 
\cite{Akhundov2, DonoghueReview, LivingReviews} and the interested reader can get more information from them.

From the usual relation between $g_{00}$ and the potential $\Phi$, $g_{00}=\left(1+\frac{2\Phi}{c^{2}}\right)$, the corrected $g_{00}$ 
component of the metric can be written as 
\begin{equation} 
\label{metric}
g_{00}=\left(1-\frac{2GM}{rc^{2}}\right)+\frac{2G^{2}M\hbar}{c^{5}}\frac{\gamma}{r^{3}}.
\end{equation}

The procedure from here on will be to look for changes in the horizon of a black hole as a consequence of eq. (\ref{metric}).
This was suggested in \cite{Duff} without giving the value of the new horizon and drawing the 
conclusion of the correction.
Let us mention that,
without summing the entire perturbation series in $\hbar$, 
we should be careful in expanding our results up to the first order in $\hbar$.
Equally we can talk with confidence about the metric only at large values of $r$ which means quantitatively
that we should respect $r \gg l_p$. Without the full quantum mechanical correction the issue of 
the central singularity remains unknown.

We assume that the corrected metric is such that 
\begin{equation} \label{grr}
-g_{00}=g^{rr} 
\end{equation}
as in the classical Schwarzschild case. Note that this assumption 
bears some non-perturbative elements. 
Indeed, it is not anymore of the form: classical result plus $\hbar$-corrections as we have it in
equation (\ref{metric}) (we will, however, expand the latter results consistently in $\hbar$ as emphasized before).
As we mentioned in the Introduction this step is motivated by the classical Schwarzschild case where equation (\ref{grr})
holds and justified a posteriori by the correct results
concerning the corrections to the Bekenstein entropy of the black holes.

We look for the values of $r$ that will make $g^{rr}=0$ 
and identify them as possible horizons for the black hole. The relevant equation reads 
\begin{equation} \label{horizon}
\left( 1 - \frac{2GM}{rc^{2}}\right)+\frac{2G^{2}M\hbar}{c^{5}}\frac{\gamma}{r^{3}}=0.
\end{equation}
It is convenient to work in dimensionless variables. To this end we introduce $y\equiv \frac{r}{r_{s}}$
with  $r_{s}=\frac{2GM}{c^{2}}$ and multiply on both sides of eq. (\ref{horizon}) with $r^{3}/r_{s}^{3}$. This way we arrive at 
\begin{equation}
y^{3}-y^{2}+\beta=0
\end{equation}
with 
\begin{equation}
\beta\equiv\frac{l_{p}^{2}}{r_{s}^{2}}\gamma . 
\end{equation}
To solve this polynomial equation we first reduce the equation by the substitution $\xi=y-\frac{1}{3}$
which results in
\begin{equation} \label{horizoneq}
\xi^{3}-\frac{1}{3}\xi-\frac{2}{27}+\beta=0.
\end{equation}
The discriminant $D$ of this cubic equation, namely
\begin{equation}
D=\left[\frac{1}{3} \left(-\frac{1}{3} \right)\right]^{3}+\left[\frac{1}{2}\left(-\frac{2}{27}+\beta \right) \right]^{2}=\frac{\beta^{2}}{4}-\frac{\beta}{27}.
\end{equation}
gives us the necessary information on the number of the real roots \cite{Bronstein}. We recall that:
\begin{itemize}
\item{If $D>0$ the polynomial in $\xi$ has only one real solution.}
\item{If $D<0$ the polynomial in $\xi$ has three real solutions.}
\item{If $D=0$ the polynomial in $\xi$ has two real solutions.}
\end{itemize}
In the following we will consider the two cases according to the sign of $\gamma$ (and hence also of $\beta$) putting
some emphasis on case when $\gamma$ is negative.

\section{Thermodynamics}

After taking $\beta=-|\beta|$, which is equivalent to having $\gamma < 0$, the reduced third order polynomial in $\xi$ can be written as
\begin{equation}
\xi^{3}-\frac{1}{3}\xi-\frac{2}{27}-|\beta|=0,
\end{equation}
and the discriminant
\begin{equation}
D=\frac{|\beta|^{2}}{4}+\frac{|\beta|}{27}
\end{equation}
is always positive definite.
Hence, only one real solution exists to the polynomial in $\xi$, i.e., $\xi_{1}$. 
The solutions, expanded in $\beta$, read
\begin{eqnarray}
\xi_{1}&=& \frac{2}{3}+|\beta|+\mathcal{O}(\beta)^{3/2}\\
\xi_{2}&=&-\frac{1}{3}+i\sqrt{|\beta|}-\frac{|\beta|}{2}+\mathcal{O}(\beta)^{3/2}\\
\xi_{3}&=&-\frac{1}{3}-i\sqrt{|\beta|}-\frac{|\beta|}{2}+\mathcal{O}(\beta)^{3/2}.
\end{eqnarray}
The horizon receives a quantum mechanical correction of the form
 \begin{equation}
 r_{nh}=r_{s}+|\beta| r_{s}=r_{s}+\frac{l_{p}^{2}}{r_{s}}|\gamma|
 \end{equation}

Within our previous assumption, $-g_{00}=g^{rr}$, the temperature of the black hole is given by $T=\frac{\hbar}{2\pi c}\kappa$, where we take $k_{B}=1$ and $\kappa$ 
is the surface gravity defined as
\begin{equation} \label{kappa} 
\kappa =\frac{1}{2} \lim_{r \to r_{nh}}\frac{\partial_{r}g_{00}}{|g_{00}g^{rr}|}.
\end{equation}
After some algebraic manipulations we find the surface gravity to be 
 \begin{equation}
 \kappa=\frac{GM}{\left(\frac{2GM}{c^{2}}+\frac{\hbar}{2Mc}|\gamma| \right)^{2}}+\frac{3G^{2}M\hbar}{c^{3}}\frac{|\gamma|}{\left(\frac{2GM}{c^{2}}+\frac{\hbar}{2Mc}|\gamma| \right)^{4}}.
 \end{equation}
The black hole temperature suitably expanded in $\hbar$ takes the simple expression  
\begin{equation}
\label{beta}
T=\frac{\hbar}{2\pi c}\left(\frac{c^{4}}{4GM}+\frac{c^{5}\hbar}{16 G^{2}M^{3}}|\gamma|+\mathcal{O}(\hbar)^{2} \right).
\end{equation}
The black hole entropy is computed using the first law of black hole thermodynamics \cite{BHthermo} $dS=c^2dM/T$. In integral form it is written as
\cite{Susskind2005}
\begin{equation}
\label{intS}
\int dS=\frac{2\pi c^{3}}{\hbar}\int \frac{dM}{\left(\frac{c^{4}}{4GM}+\frac{c^{5}\hbar}{16 G^{2}M^{3}}|\gamma|+\mathcal{O}(\hbar)^{2} \right)}.
\end{equation}
We left here the limits of integration open as they depend whether or not we have a black hole remnant which we will discuss below. 
After expanding the expression inside the integral to first order in $\hbar$,
\begin{equation}
S=\frac{2\pi c^{3}}{\hbar}\int dM \left[\frac{4GM}{c^{4}}-\frac{\hbar}{c^{3}M}|\gamma| + \mathcal{O}(\hbar^{2})  \right],
\end{equation}
and making the substitution
\begin{equation}
\chi= \frac{M}{m_{p}}=M\sqrt{\frac{G}{\hbar c}},
\end{equation}
we obtain
\begin{eqnarray}
S&=&2\pi \int d\chi \left[ 4\chi -\frac{|\gamma|}{\chi}+\mathcal{O}(\hbar)^{2}\right]\\ \nonumber
&=&4\pi \chi^{2}-2\pi |\gamma| \ln [\chi]+\mathcal{O}(\hbar)^{2}.
\end{eqnarray}
Going back to our usual variables we get the final result
\begin{equation}
S=\frac{4\pi G M^{2}}{\hbar c}k_{B}-2\pi |\gamma| k_{B} \ln \left[ M \sqrt{\frac{G}{\hbar c}} \right]+\mathcal{O}(\hbar)^{2}.
\end{equation}

\begin{equation}
\label{entropy}
S=S_{BH}-\pi |\gamma | \ln\left[ \frac{\mathcal{A}}{4l_{p}^{2}} \right]+\pi |\gamma| \ln\left[4\pi \right]+\mathcal{O}(\hbar)^{2},
\end{equation}
where we introduced the Schwarzschild black hole area, i.e. $\mathcal{A}=4\pi r_{s}^{2}=\frac{16 \pi G^{2}M^{2}}{c^{4}}$ and 
the classical expression obtained by Bekenstein \cite{bekenstein} is
$S_{BH}=\mathcal{A}/4l^2_p$.

Furthermore, Eq. (\ref{intS}) has an exact solution given by
\begin{eqnarray}
\label{entropy1}
S&=&S_{BH}-\pi |\gamma | \ln\left[ \frac{\mathcal{A}}{4l_{p}^{2}}\frac{1}{\pi}+|\gamma| \right]+\mathcal{O}(\hbar)^{2} \nonumber \\
&\approx&S_{BH}-\pi |\gamma | \ln\frac{\mathcal{A}}{4l_{p}^{2}}+ \pi |\gamma|\ln \pi 
+\mathcal{O}(\hbar)^{2}.
\end{eqnarray}

We will comment on this result in section VI, but we note already here that 
there exists an overwhelming agreement in the literature on 
logarithmic corrections to the
black hole entropy \cite{Kaul2000,Medved2004,Camelia2004,Meissner2004,Das2002,Domagala2004,Chatterjee2004,Akbar2004,Myung2004,Chatterjee2005,Carlip2000,Ghosh2005,Fursaev1995,Nojiri2003,Kaul2012,Hod2004,Hai-Xia2007,Tawfik2015,Mukherji2002,Wang2008,Banerjee2008}. Whereas most of the results use a model for Quantum Gravity, we have
obtained corrections of the same form by analyzing $\hbar$ corrections to the Newtonian
potential via an effective theory of gravity.

\section{Heat Capacity and the Black Hole Remnant}
Let us compute the heat capacity of the black hole using the standard expressions $C=c^{2}\frac{d M}{d T}$. 
From eq. (\ref{beta}) we can expand $T^{-1}$ up to order one in $\hbar$ and deduce two solutions
for $M$ as a function of $T^{-1}$:
\begin{equation}
M_{\pm}=\frac{\frac{1}{T}\pm\sqrt{\left(\frac{1}{ T}\right)^{2}+64\frac{\pi^{2}G}{\hbar c^{5}}|\gamma|}}{\frac{16\pi G}{\hbar c^{3}}}.
\end{equation}
We note that taking $\gamma=0$ forces us to consider the positive sign in order to recover the usual case for 
Schwarzschild black hole. Therefore, the
heat capacity turns out to be
\begin{equation}
\label{heatc}
C=-\frac{2\pi}{m_{p}^{2}}\left[ \frac{ (-4M^{2}+m_{p}^{2}|\gamma|)^{2}}{4M^{2}+m_{p}^{2}|\gamma|}\right].
\end{equation}

At this point, let us define the remnant mass, $M_{r}$, by $C(M_r)=0$ \cite{Banerjee2010,Dutta2014,Gangopadhay2015,Xiang2007,Xiang2009}. 
Thus, when $M_r$ is reached, the black hole evaporation stops. From eq. (\ref{heatc}) we obtain
\begin{equation}
M_{r}=\frac{\sqrt{|\gamma|}}{2}m_p
\end{equation}
which is of the order of the Planck mass $m_p$.
We can relate this remnant mass to a maximum temperature, by taking
eq. (\ref{beta}) and replacing the mass by the value of the remnant mass, $M_r$. This yields:
\begin{eqnarray}
T(M_{r})\equiv T_{max}&=&\frac{\hbar^{1/2}c^{5/2}}{4\pi G^{1/2}}\left[ \frac{1}{|\gamma|^{1/2}}+\frac{1}{|\gamma|^{3/2}}\right] \\ \nonumber
&=&\frac{c^{2}}{4\pi}m_{p}\left[\frac{1}{|\gamma|^{1/2}}+\frac{1}{|\gamma|^{3/2}} \right]
\end{eqnarray}
which is of the order of $T_{0}=c^{2}m_{p}$, a number suggested by Sakharov \cite{Sakharov} for the the maximum 
temperature of thermal radiation. We have based the black hole remnant on the zero value of the heat capacity.
In a different context a black hole remnant emerges when the temperature becomes complex \cite{Adler1,Adler2001} where the maximum temperature
is of the same order as here. 

As we have already mentioned the validity of the approach might break down at the Planck scales due to the lack
of unitarity of the amplitude at these scales. 
Higher order correction might become important here.
Therefore the above results are to be taken with caution.
We have included them because the one-loop result often shows the right tendency of the final phenomenological effect.
Secondly, as mentioned before equation (\ref{grr}) is essentially non-pertubative and should, in principle, take some
of the higher order effects into account.   
Indeed, both, the remnant black hole mass of the order of the Planck mass and the maximum temperature have analogies in the literature.
We point out that the existence of a remnant in the $\gamma<0$ case is in complete agreement with similar
conclusions obtained within the quadratic GUP formalism \cite{Banerjee2010,Dutta2014}. Hawking radiation formulated within the 
formalism of a generalized uncertainty relation also indicates a black hole remnant as shown in \cite{Adler1}.
Including the cosmological constant $\Lambda$ the generalized uncertainty relation not only gives the maximum temperature and
minimum mass, but in addition also a minimum temperature of the order of $\sqrt{\Lambda}$ and a maximum mass proportional $m_p^2/\sqrt{\Lambda}$
\cite{GUPL}.  
 
\section{The hypothetical case $\gamma > 0$}
We follow throughout the paper the latest state-of-art and  consider 
the case of negative $\gamma$, to be precise, we take $\gamma =-41/10$ as the correct value.
Nevertheless, it is illustrative 
to demonstrate how the physics changes when going from a negative $\gamma$ to a positive result.
We will discuss below some salient features of the sign of the parameter $\gamma$. Indeed, as it will turn out
it is the sign which changes the most important physical aspects. 

For $\gamma > 0$ we simply have
\begin{eqnarray}
D&=&\frac{\beta^{2}}{4}-\frac{\beta}{27}>0 \mbox{ if } \beta>\frac{4}{27}\\
D&<&0 \mbox{ if } \beta<\frac{4}{27}\\
D&=&0 \mbox{ if } \beta=\frac{4}{27}.
\end{eqnarray}
The  value $\beta_{1}=\frac{4}{27}\simeq0.14848$ is the benchmark which decides the number of horizons in this case. 
Before going into the details of the relevant physical aspects let us complete the 
the solutions for the horizons. Applying standard prescriptions one parametrizes the solutions by
\begin{equation}
u=\left[ \frac{1}{27}-\frac{\beta}{2}+\sqrt{D}\right]^{1/3} \mbox{ ; } v=\frac{1}{9}\left[ \frac{1}{27}-\frac{\beta}{2}+\sqrt{D}\right]^{-1/3}.
\end{equation}
The three solutions to the polynomial equation (\ref{horizoneq}) are then calculated to give
\begin{eqnarray}
\xi_{1}&=& u+v, \\
\xi_{2}&=& -\frac{u+v}{2}+\frac{u-v}{2}i\sqrt{3},\\
\xi_{3}&=&-\frac{u+v}{2}-\frac{u-v}{2}i\sqrt{3}.
\end{eqnarray}
The physics becomes interesting if we consider $\beta<<1$. 
Indeed, for the opposite case we cannot be sure if we are still within range of the validity of our calculations.
Three real solutions emerge now. 
Going back to the expressions in $y$ ($y=\xi+\frac{1}{3}$ and recalling that $y=r/r_s$) we obtain after expanding in
$\beta$
\begin{eqnarray}
y_{1}&=&1-\beta+\mathcal{O}(\beta)^{3/2},\\
y_{2}&=&-\sqrt{\beta}+\frac{\beta}{2}+\mathcal{O}(\beta)^{3/2},\\
y_{3}&=&\sqrt{\beta}+\frac{\beta}{2}+\mathcal{O}(\beta)^{3/2}.
\end{eqnarray}
We note here that $y_{2}$ can be discarded since it is negative for small values of $\beta$. Specifically,
\begin{equation}
y_{2}=-\sqrt{\beta}+\frac{\beta}{2}>0 \mbox{ if } \beta>4.
\end{equation}
Summarizing, we arrive at a quantum correction to the standard horizon, i.e., 
\begin{equation}
r_{nh+}=r_{s}-\beta r_{s}=r_{s}-\frac{l_{p}^{2}}{r_{s}}\gamma
\end{equation}
and, interestingly, at a new purely quantum-mechanical horizon, which vanishes in the limit
$\hbar \to 0$,  of the form
\begin{equation}
r_{nh-}=\sqrt{\beta}r_{s}+\frac{\beta r_{s}}{2}=l_{p}\sqrt{\gamma}+\frac{l_{p}^{2}}{2r_{s}}\gamma . 
\end{equation}
It is remarkable that quantum mechanical corrections would reveal the existence of a new horizon (disregarding the
fact that we are discussing here a hypothetical case of $\gamma$).
In theories inspired by non-commutative geometry \cite{noncomm} a similar phenomenon occurs.

We saw that if $\gamma>0$ we have two horizons provided $\beta<<1$.
In this case the surface gravity is calculated taking only the outermost horizon into account.
The relevant expression is
\begin{equation}
\kappa =\frac{1}{2} \lim_{r \to r_{+}}\frac{\partial_{r}g_{00}}{|g_{00}g^{rr}|}
\end{equation}
where $r_{nh+}=r_{+}$ and $r_{nh-}=r_{-}$. Therefore. the full surface gravity reads as
\begin{equation}
\kappa=\frac{GM}{\left(\frac{2GM}{c^{2}}-\frac{\hbar}{2Mc}|\gamma| \right)^{2}}-\frac{3G^{2}M\hbar}{c^{3}}\frac{|\gamma|}{\left(\frac{2GM}{c^{2}}-\frac{\hbar}{2Mc}|\gamma| \right)^{4}}.
\end{equation}
One can easily calculate the temperature which turns out to be
\begin{equation}
\label{positivegammatemp}
T=\frac{\hbar}{2\pi c}\left(\frac{c^{4}}{4GM}-\frac{c^{5}\hbar}{16 G^{2}M^{3}}|\gamma|+\mathcal{O}(\hbar^{2}) \right).
\end{equation}
Finally, the entropy can be computed along the same lines as in the
case $\gamma < 0$ and now takes the form
\begin{equation}
S=S_{BH}+\pi |\gamma | \ln\left[ \frac{\mathcal{A}}{4l_{p}^{2}} \right]-\pi |\gamma| \ln\left[4\pi \right]+\mathcal{O}(\hbar)^{2}
\end{equation}
or
\begin{eqnarray}
\label{entropybis}
S&=&S_{BH}+\pi |\gamma | \ln\left[ \frac{\mathcal{A}}{4l_{p}^{2}}\frac{1}{\pi}+|\gamma| \right]+\mathcal{O}(\hbar)^{2} \nonumber \\
&\approx&S_{BH}+\pi |\gamma | \ln\frac{\mathcal{A}}{4l_{p}^{2}}- \pi |\gamma|\ln \pi 
+\mathcal{O}(\hbar)^{2}
\end{eqnarray}
if the integral given by Eq. (\ref{intS}) is solved exactly, as in the $\gamma < 0$ case.
Notice that in this hypothetical case we obtain a different relative sign of the correction
to the black hole entropy.

To make the analysis of the case $\gamma > 0$ complete, it remains to 
inspect the possibility of a black hole remnant.
After expanding $1/T$ in eq. (\ref{positivegammatemp}) up to order $\hbar$,
we find again
two solutions for $M$ as a function of $T^{-1}$. They read
\begin{equation}
M_{\pm}=\frac{\frac{1}{T}\pm\sqrt{\left(\frac{1}{ T}\right)^{2}-64\frac{\pi^{2}G}{\hbar c^{5}}|\gamma|}}{\frac{16\pi G}{\hbar c^{3}}}
\end{equation}
As before, to recover the standard case we choose the positive sign in this equation. 
The heat capacity is easily calculable to be 
\begin{widetext}
\begin{equation}
C=-\frac{\pi\left(4M^{2}+m_{p}^{2} \gamma \right)^{2}\left(m_{p}^{2}\gamma+4M^{2}+\left|m_{p}^{2}\gamma-4M^{2}\right|  \right)}{4 M^{2}m_{p}^{2} \left |m_{p}^{2}\gamma -4M^{2}\right|}.
\end{equation}
\end{widetext}
It is clear that no remnant mass shows up in this case, i.e. there does not exist a value for $M$ which makes $C=0$.
Indeed, the $\gamma>0$ case seems to be  more subtle. In this situation, 
the logarithmic correction acquires the opposite sign compared to that predicted by GUP and others.

\section{Logarithmic Corrections to the Black Hole Entropy in Different Models}

Different approaches to quantum gravity  have predicted corrections to the Bekenstein--Hawking entropy for black holes of
the form \cite{Kaul2000,Medved2004,Camelia2004,Meissner2004,Das2002,Domagala2004,Chatterjee2004,Akbar2004,Myung2004,Chatterjee2005}
\begin{equation}
\label{eqfirst}
S=\frac{\mathcal{A}}{4 l_{p}^{2}}+c_{0}\ln \left(\frac{\mathcal{A}}{4 l_{p}^{2}} \right)+\sum_{n=1}^{\infty}c_{n}\left(\frac{\mathcal{A}}{4l_{p}^{2}} \right)^{-n},
\end{equation}
where the $c_{n}$ coefficients are parameters which depend on the specific model considered. 
Interestingly, loop quantum gravity calculations are used to fix 
$c_{0}=-1/2$ \cite{Meissner2004}. Moreover, the deformed commutation relations giving place to a generalized uncertainty principle (GUP) 
have been also used to compute the effects of the GUP on the black hole entropy from different perspectives (see, for example, 
\cite{Medved2004,Adler2001,Camelia2006,Majumder2011,Majumder2013}). In this case the expression for the entropy reads
\begin{equation}
\label{GUP}
S= \frac{\mathcal{A}}{4 l_{p}^{2}}+\frac{\sqrt{\pi} \alpha_{0}}{4}\sqrt{\frac{\mathcal{A}}{4l_{p}^{2}}}-\frac{\pi \alpha_{0}^{2}}{64}\ln 
\left(\frac{\mathcal{A}}{4 l_{p}^{2}} \right)+
\mathcal{O}(l_{p}^{3}),
\end{equation}
where $\alpha_{0}$ measures the deviation for the standard Heisenberg case, i.e.,
\begin{equation}
x_{i}=x_{0i}\; ;  \; p_{i}=p_{0i}\,\left(1-\alpha p_{0}+ 2 \alpha^{2} p_{0}^{2}\right),
\end{equation}
where $\left[x_{0i},p_{0j}\right]=i \hbar \delta_{ij}$ and $p_{0}^{2}=\sum_{j=1}^{3}p_{0j}p_{0j}$ and
$\alpha = \nobreak \alpha_{0}/m_{p}c$, with being $\alpha_{0}$ a dimensionless constant \cite{Ali2009,Ali2011}. 
Even more, polymerization (a non--standard representation of quantum mechanics that was inspired by loop quantum
gravity, LQG) also predicts logarithmic corrections to the black hole entropy \cite{Gorji2014} 
(it has been shown that polymerization and quadratic GUP are equivalent
provided $\alpha_{0}$ and the polymerization parameter are proportional \cite{Pedro2015}). In fact, the leading--order corrections to the entropy of any 
thermodynamic system due to small statistical fluctuations around equilibrium, when applied to black holes, are shown to be of the form $\ln[\mathcal{A}]$
\cite{Das2002}. Of course, the four dimensional Schwarzschild BH has a negative
heat capacity and to apply the previous idea, the authors of \cite{Das2002} considered an AdS--Schwarzschild BH and take the limit of large $\Lambda$ to account for a
stable system.  Once this is realized, the leading corrections are shown to be logarithmic. In this sense, the authors of \cite{Das2002} claim that log--corrections are
universal.

Therefore, the corrections given by eq. (\ref{entropy}), obtained from one--loop calculations, are consistent with different approaches which incorporate, in some sense, some quantum gravitational considerations. Specifically, the correct sign for the logarithmic
term is obtained for the $\gamma<0$ case. As we commented before, this is the case of LQG and of quadratic GUP. Therefore, our
approach is consistent with both of them provided $|\gamma|=(2 \pi)^{-1}=\alpha_{0}^{2}/64$. Although the $\gamma>0$ case
predicts the existence of a second horizon, this case corresponds also to linear GUP
provided $\gamma=\alpha_{0}^{2}/64$, as in the previous case. Interestingly, we note that, in a recent work \cite{Elias2017}, the authors
propose a technique to compute the deformation parameter of the GUP by using the leading quantum corrections to the Newtonian potential, somehow in the
same spirit of our work.

We would like to note that our analysis uses macroscopic black hole properties related by the first law. In this sense, no use of properties of
the underlying microscopical theory, whatever it will be, are used. Although we are aware of the thermodynamical instability of the Schwarzschild 
black hole without putting it in a box, we expect that the case here considered turn to be thermodynamically stable by adding a cosmological constant. Therefore,
following \cite{Das2002}, the corrections to the canonical entropy due to small statistical fluctuations around equilibrium will be 
shown to be of the form $\ln[\mathcal{A}]$. Even more, although the coefficient of the logarithmic corrections (and even their sign) 
depend on the particular scheme one used to define the entropy \cite{Page2005}, the logarithmic dependence seems to be universal.


In this sense, we think that our comparison with other models will not be complete unless Euclidean path integral methods are taken into account 
\cite{Gibbons1977}. 
As is well known,  in the stationary phase and one--loop approximation, the leading correction to the partition function is given by the metric which is an
Euclidean classical solution and the one--loop contribution comes from the action due to matter fields. In this sense, the physical interpretation of the
logarithmic corrections in this approach are clear: they come from the spectrum of massless fields and their coupling to the gravitational background.

In particular, the first Euclidean gravity calculation on the log--correction for a Schwarzschild BH has been recently reported by A. Sen
\cite{Sen2013}. Let us recall that the exact numerical pref-actor accompanying the log--correction depends on the employed ensemble and on the number
of fields (other than gravity) included in the calculation. However, the log--correction
is recovered in all cases. Moreover, as pointed out by Sen, {\it the macroscopic results for logarithmic corrections seem quite robust} since loops
(higher than one) do not give any log--correction.  

We are aware that the statistical interpretation provided for the log--correction (which arise in this context even without matter simply by graviton
fluctuations) by the path integral method 
is not obvious in our approach. 
However, although the particular numerical pref-actor of these approaches does not agree in general (the
same lack of agreement is also present in the path integral and loop quantum gravity comparison),
we think that in our work we show explicitly the universality of log--corrections to BH entropy from an effective field theory approach.
This idea can be traced back to a recent work by Bjerrum--Bohr, Donoghue and Vanhove \cite{Bohr2014}. Specifically, as recently expressed by 
the authors of this work, {\it we stress that our results are universal and thus will hold in any quantum theory of gravity with the same
low--energy degrees of freedom as we are considering}. Therefore, both the quantum correction to the Newtonian potential, the corresponding
correction to the Schwarzschild metric \cite{DonoghueReview} and the logarithmic correction to the entropy are among these universal results.

In retrospect, the agreement with other findings on the corrections to the black hole entropy gives us some confidence about the quantum mechanical
corrections to the Newtonian potential and the conclusions drawn from it. 

Finally, we note that there exist other approaches to the quantum aspects of the Schwarzschild metric and we refer the reader
to the original literature (see \cite{Lake} and references therein).

\section{The geodesic equation of motion}
One of the key observables in General Relativity is, of course, the particle trajectory once the metric in which the particle moves has been
given. 
From our point of view, it is crucial to check whether the quantum mechanical corrections proposed above will
change the standard predictions in a drastic way. This would be the case
if, for instance,  new circular (stable or unstable) orbits would appear leading to new phenomenological results for the particle trajectory. 
It is worthwhile to note that even if the corrections to the geodesic equation of motion come out to be proportional to $l_p$,
there is no a priori guarantee that all observables will receive small corrections and that no new 
features will emerge. Small quantum effects on the three body Lagrangian points were recently found using the
same corrections to the Newtonian potential \cite{Batista}. 
Mixing of scales can lead to new results as it happens, e.g., in the Schwarzschild-de Sitter metric
where scales of the cosmological constant combined with the Schwarzschild radius reveal new
aspects of the effective potential \cite{scalesLambda}.

\subsection{The quantum corrections to the effective potential}
As far as the metric is concerned the results of the previous sections can be summarized by writing 
\begin{equation} \label{metric2}
ds^{2}=B(r)dt^{2}-A(r)dr^{2}-r^{2}C(r)[d\theta^{2}+\sin^{2}\theta d\phi^{2}]
\end{equation}
where $C(r)=1$, $A(r)=\frac{1}{B(r)}$ and 
\begin{equation}
B(r)=\left(1-\frac{r_{s}}{r}+\frac{r_{s}l_{p}^{2}}{r^{3}}\gamma\right).
\end{equation}
By using the standard methods,  
one can cast the equation of motion in the form
\begin{equation}
\frac{\dot{r}^{2}}{2}+V_{eff}(r)=\mbox{const}
\end{equation}
in which the effective potential can be split into two terms indicating the classical and the quantum part
 \begin{equation} \label{Veff}
V_{eff}(r)=V_{eff}^{(\hbar^{0})}(r)+V_{eff}^{(\hbar)}(r). 
\end{equation}
The respective contributions read
\begin{equation}
V_{eff}^{(\hbar^{0})}(r)=
\begin{cases}
-\frac{GM}{r}+\frac{l^{2}}{2r^{2}}-\frac{GMl^{2}}{c^{2}r^{3}} & \text{if}\ m\neq0\\
\frac{l^{2}}{2r^{2}}-\frac{GM l^{2}}{c^{2}r^{3}} & \text{if}\ m=0
\end{cases}
\end{equation}
and
\begin{equation}
V_{eff}^{(\hbar)}(r)=
\begin{cases}
\frac{G^{2}M\hbar}{c^{3}r^{3}}\gamma+\frac{G^{2}M l^{2} \hbar}{c^{5}r^{5}}\gamma & \text{if}\ m\neq0\\
\frac{G^{2}M l^{2}\hbar}{c^{5}r^{5}}\gamma & \text{if}\ m=0
\end{cases}
\end{equation}
with $m$ the mass of the test particle and $l$ the angular momentum per mass.
Evidently, the quantum correction vanishes as $\hbar \rightarrow 0$.
In the following
we will study the extrema and zeros of the corrected effective potential. 
If no new zeros and local extrema emerge as a result of the quantum corrections 
and the new zeros and extrema receive corrections of the order of $\hbar$, we can consider the
theory based on (\ref{metric2}) as consistent and in accordance with observational facts. In the antipodal case of additional extrema,
even if as a small effect, would imply new stable and unstable circular orbits.
\subsection{The massless case}
Based on
\begin{equation}
V_{eff}(r)=\frac{l^{2}}{2}\left(\frac{1}{r^{2}}-\frac{r_{s}}{r^{3}}+\frac{l_{p}^{2}r_{s}}{r^{5}}\gamma \right)
\end{equation}
we look first for the zeros ($r_{0}$) of this function. With $r_{0}\neq0$ we obtain
\begin{equation}
r_{0}^{3}-r_{s}r_{0}^{2}+l_{p}^{2}r_{s}\gamma=0
\end{equation}
Dividing by $r_{s}^{3}$ and defining $x\equiv \frac{r_{0}}{r_{s}}$, $\beta\equiv \frac{l_{p}^{2}}{r_{s}^{2}}\gamma$ 
and  setting $y=x-\frac{1}{3}$ we arrive at a third
order polynomial whose zeros we wish to find,  i.e., 
\begin{equation}
y^{3}-\frac{1}{3}y-\frac{2}{27}+\beta=0
\end{equation}
As usual it is the discriminant of this equation which is of importance. The latter is given by \cite{Bronstein}
\begin{equation}
D=\frac{1}{4}\beta^{2}-\frac{1}{27}\beta.
\end{equation}
The case of relevance turns out to be  $D > 0$ which implies one real solution of the cubic polynomial. The reason 
is that
\begin{equation}
D>0 \mbox{ implies } \beta<0 \mbox{ or } 0<\beta<\frac{4}{27},
\end{equation}
or, equivalently,
\begin{equation}
D>0 \mbox{ if } \gamma <0 \mbox{ or } \gamma>0 \mbox{ and } r_{s}^{2}<\frac{27}{4}l_{p}^{2}\gamma.
\end{equation}
On the other hand, the case with three real solutions would lead to
\begin{equation}
D<0 \mbox{ if } \gamma >0 \mbox{ and } r_{s}^{2}>\frac{27}{4}l_{p}^{2}\gamma.
\end{equation}
Again it is the sign of $\gamma$ which is crucial here. Since we decided to
focus on the negative value of $\gamma$, it suffices to handle the case
$D >0$. 
The only real zero is then calculated to be
\begin{equation}
r_{0}=r_{s}-\frac{l_{p}^{2}}{r_{s}}\gamma +\mathcal{O}(l_{p})^{3}.
\end{equation}

The method to find the extreme goes, in principle, along the same lines as outlined above. Putting the derivative of the 
effective potential to zero results in a third order equation in $r_{max}$. The latter can be 
transformed in a third order equation in the variable $\xi=(r_{max}/r_{s})-\frac{1}{2}$
\begin{equation}
\xi^{3}-\frac{3}{4}\xi-\frac{1}{4}+\frac{5}{2}\beta=0.
\end{equation}
The discriminant in this case can be calculated to be
\begin{equation}
D_{max}=\frac{25}{16}\beta^{2}-\frac{5}{16}\beta.
\end{equation}
The distinct cases for $D_{max}$  are similar to the discriminant of the zeros discussed above.
In short, we can summarize it as follows 
\begin{eqnarray}
&D_{max}>0& \mbox{ if } \left(\beta<0 \mbox{ or } \beta >\frac{1}{5}\right)\Rightarrow \left( \gamma<0 \mbox{ or } r_{s}^{2}<5l_{p}^{2}\gamma\right) \\
&D_{max}<0& \mbox{ if } 0<\beta<\frac{1}{5} \Rightarrow \gamma>0 \mbox{ and } r_{s}^{2}>5l_{p}^{2}\gamma \\
&D_{max}=0& \mbox{ if } \gamma>0 \mbox{ and } r_{s}^{2}=5l_{p}^{2}\gamma
\end{eqnarray}
Continuing with $\gamma < 0$ we find again only one real zero. The radius of the unstable photon circular
orbit receives a small correction proportional to $\hbar$. In particular, we obtain
\begin{equation}
r_{max}=\frac{3}{2}r_{s}-\frac{10}{9}\frac{l_{p}^{2}}{r_{s}}\gamma + \mathcal{O}(l_{p})^{3}.
\end{equation}
The first term in $r_{max}$ is the standard result of the effective potential without quantum corrections.

\subsection{The massive case}
For the massive case we write the effective potential as follows
\begin{equation}
V_{eff}(r)=\frac{l^{2}}{2}\left(\frac{1}{r^{2}}-\frac{r_{s}}{r^{3}}+\frac{r_{s}l_{p}^{2}}{r^{5}}\gamma \right) -\frac{r_{s}}{2r}c^{2}+\frac{r_{s}l_{p}^{2}}{2r^{3}}\gamma c^{2}.
\end{equation}
The search for the zeros as well as the extrema gives now a fourth order polynomial equation.
Even though the latter can be solved analytically the calculations are quite extensive and the steps
not illuminating. We skip all the details and quote the final result obtained with the help
of MATHEMATICA. The physically relevant zeros, i.e., the zeros of the effective potential which lie outside
the horizon are 
\begin{equation}
r_{0}^{1,2}=\frac{l^{2}}{2c^{2}r_{s}}\mp \frac{1}{2}\sqrt{\frac{l^{4}}{c^{4}r_{s}^{2}}-4\frac{l^{2}}{c^{2}}}-\frac{1}{2}\left(1\pm \frac{\frac{l^{2}}{c^{2}r_{s}}}{\sqrt{\frac{l^{4}}{c^{4}r_{s}^{2}}}-4\frac{l^{2}}{c^{2}}} \right)\frac{l_{p}^{2}}{r_{s}}\gamma+\mathcal{O}(l_{p}^{4})
\end{equation}
As $\hbar \rightarrow 0$ we recover the usual roots.
The extrema are located at
\begin{equation}
r_{max}^{1,2}=\frac{l^{2}}{c^{2}r_{s}}\mp \sqrt{\frac{l^{4}}{c^{4}r_{s}^{2}}-3\frac{l^{2}}{c^{2}}}-\frac{1}{9}\left( 5\pm \frac{12r_{s}+10\frac{l^{2}}{c^{2}r_{s}}}{\sqrt{\frac{l^{4}}{c^{4}r_{s}^{2}}}-3\frac{l^{2}}{c^{2}}}\right)\frac{l_{p}^{2}\gamma}{r_{s}}.
\end{equation}
The restrictions to make both results real are actually restrictions on $l^{2}$  and are also present in the standard general
relativistic case without quantum corrections. 
For the $r_{0}$ values to exist we have to impose $\frac{l^{2}}{c^{2}}>4r_{s}^{2}$ and, for the $r_{max}$ values to exist, 
the inequality to be satisfied is $\frac{l^{2}}{c^{2}}>3r_{s}^{2}$. This stronger case does not appear accidentally here.
In the classical case of the Schwarzschild metric, only respecting $l < l_{crit}=\sqrt{3}r_{s}c$ leads to an effective potential
without local extrema.  
  
\section{Conclusions}
We have explored the consequences of quantum mechanical corrections to the Newtonian potential.
This correction in tandem with $-g_{00}=g^{rr}$ fixes the metric.
We probe into the physics around the
horizon of this metric. We find
a corrected Schwarzschild horizon where the correction is proportional to $\hbar$.
This was used to infer the corrections to the black hole entropy. We derived
logarithmic corrections in agreement with many other approaches. A black hole
remnant of the order of Planck mass emerges in this case. 
The (hypothetical) positive correction
to the Newtonian potential gives another picture. In addition to the quantum
mechanically corrected Schwarzschild radius, a second horizon of purely quantum
mechanical nature (proportional to $\sqrt{\hbar}$ and $\hbar$) is possible. 
No remnant
exists in this case. 
It is obvious how much the sign of the correction affects the conclusions.
Finally, we examine the consequences of the $\hbar$ correction in the geodesic
equation of motion and find that that classical tests of General Relativity will
be affected only marginally.

In conclusion, 
the simple quantum mechanical correction to the Newtonian
potential taken together with a reasonable assumption on the $g^{rr}$ component has remarkable consequences. Whether Hawking radiation or Bekenstein entropy
the quantum mechanics in the gravity of a black hole is focused at the horizon.
We added to this list a quantum mechanical correction of the horizon and
connected it with the correction to the entropy.

\section*{Acknowledgment}
P. B. and M. N. acknowledge the support from the Faculty of Science and Vicerector\'{\i}a de Investigaciones of
Universidad de Los Andes, Bogot\'a, Colombia. M. N. and S. B. M. would like to thank 
the  administrative  department  of  science,  technology  and  innovation of Colombia (COLCIENCIAS) 
for  the financial support provided. We also thank E. C. Vagenas for useful discussions.

\end{document}